\documentclass[12pt]{article}
\usepackage[english]{babel}
\usepackage{amsfonts}
\usepackage{amsmath}
\usepackage{amssymb}
\usepackage{mathrsfs}
\usepackage{latexsym}
\usepackage{multicol}
\usepackage{fancybox}
\usepackage{dsfont}
\usepackage{color}
\usepackage{hyperref}
\usepackage{subfigure} 
\usepackage{graphicx} 
\usepackage{caption}

\def\be{\begin{equation}}
\def\ee{\end{equation}}
\def\tl{\tilde} 
 
\def\gm{\gamma} 
\def\lm{\lambda}

\newtheorem{th1}{Theorem}

\newtheorem{remark}{Remark}

\def\be{\begin{equation}}
\def\ee{\end{equation}}
\def\bea{\begin{eqnarray}}
\def\eea{\end{eqnarray}}
\def\FF{\mathcal{F}}

\def\dblone{\hbox{$1\hskip -1.2pt\vrule depth 0pt height 1.6ex width 0.7pt \vrule depth 0pt height 0.3pt width 0.12em$}}

\begin{document}

\begin{center}
\Large{\bf{Quantum B\"acklund Transformations: some ideas and examples \footnote{``This paper is a
contribution to \emph{Solitons in 1+1 and 2+1 dimensions. DS, KP and all that}, conference in honor of the 70th birthday of Marco Boiti and Flora Pempinelli, Lecce, 2011 September 13-14.''}}}
\end{center}

\begin{center}
{ {\bf Orlando Ragnisco$^\dag$,  Federico Zullo$\ddag$}}

{$\dag\ddag$Dipartimento di Fisica,   Universit\`a di Roma Tre \\ $\dag$Istituto Nazionale di
Fisica Nucleare, sezione di Roma Tre\\  Via Vasca Navale 84,  00146 Roma, Italy  \\
~~E-mail: ragnisco@fis.uniroma3.it, zullo@fis.uniroma3.it}

\end{center}

\medskip
\medskip

\begin{abstract}
\noindent
In this work we give a mechanical (Hamiltonian) interpretation of the so called \emph{spectrality property} introduced by Sklyanin and Kuznetsov in the context of  B\"acklund transformations (BTs) for finite dimensional integrable systems. The property turns out to be deeply connected with the Hamilton-Jacobi separation of variables and can lead to the \emph{explicit} integration of the underlying model through the expression of the BTs. Once such construction is given, it is shown, in a simple example, that it is possible to interpret the Baxter \emph{Q} operator defining the quantum BTs us the Green's function, or propagator, of the time dependent Schr\"odinger equation for the interpolating Hamiltonian.  
\end{abstract}

\bigskip\bigskip

\noindent


\section{Introduction} \label{sec1}
Starting from the last decade of the previous millennium, a number of results on discretization of finite dimensional integrable systems appeared:  we quote for instance
the Euler top \cite{BLS}, the Lagrange top \cite{BS}, the rational Gaudin
magnet \cite{HKR} the Ruijsenaars-Schneider model \cite{NRK}, the Henon-Heiles, Garnier and Neumann
systems \cite{R}, \cite{RS}, \cite{HKR1} and others (see also \cite{Sbook} and the references
therein). It turned out \cite{KV} that  all the exact  discretizations for these systems associate
new solutions to a given one: they are BTs for such systems. These developments suggested to Sklyanin and Kuznetsov \cite{KS} that the concept of B\"acklund transformations should be revised in order to enlighten some new (and old)
aspects of the subject. Actually,  these two authors had previously obtained  fundamental results about  "separation of variables" and their  connection with the new techniques of classical and
quantum inverse scattering method.  These findings paved the way to a better understanding of 
the role of BTs in the context of finite dimensional integrable
systems and to unveiling  their deep and fruitful  links with Hamiltonian dynamics and separation of variables (for a characterization of the BTs for finite dimensional integrable systems by a geometrical point of view see \cite{KV}). Let us briefly remember some of these findings. Assume that the dynamical system under consideration is defined by a Lax pair with a spectral parameter, say $L(\lm)$ and $M(\lm)$. The trace of the powers of $L(\lm)$ will be conserved quantities, so also the determinant of the Lax matrix will be conserved. The BTs, in the case of finite dimensional systems, are \emph{canonical transformations preserving the algebraic form of the integrals}, so if $(p_i,q_i)_{i=1}^{n}$ is the set of the dynamical variables, one has: 
\be
\big(p_i,q_i\big)_{i=1}^{n}\overset{BT}{\Longrightarrow} \big(\tl{p}_i,\tl{q}_i\big)_{i=1}^{n}\qquad L(\lm, p_k, q_k)\overset{BT}{\Longrightarrow} \tl{L}(\lm)=L(\lm, \tl{p}_k, \tl{q}_k)
\ee
\be
\{\tl{p}_i,\tl{p}_j\}=\{p_i,p_j\};\;\; \{\tl{q}_i,\tl{q}_j\}=\{q_i,q_j\};\;\;\{\tl{p}_i,\tl{q}_j\}=\{p_i,q_j\}=\delta_{ij}.
\ee
Since the algebraic form of the integrals is preserved, also the determinant of the Lax matrices $L(\lm)$ and $\tl{L}(\lm)$ is the same:
$$
H_{i}(p_k,q_k)=\tl{H}_{i}(\tl{p}_k,\tl{q}_k) \;\Longrightarrow \det(L(\lm))=\det(\tl{L}(\lm))
$$
Hence,   the existence of a BTs for the given system entails that  the two Lax matrices $L(\lm)$ and $\tl{L}(\lm)$ are connected by a similarity transformation,  provided by the so called dressing matrix $D(\lm)$:
\be\label{LD=DL}
\tl{L}(\lm)D(\lm)=D(\lm)L(\lm)
\ee
Obviously the dressing matrix is not unique,  because one can have different BTs for the same system. Furthermore, if the transformations are not explicit, $D(\lm)$ might depend on both the sets $(p_i,q_i)_{i=1}^{n}$ and $(\tl{p}_i,\tl{q}_i)_{i=1}^{n}$ More interesting are parametric and explicit BTs, that is BTs depending on one or more  parameters,  arbitrary (or restricted to some range of values) such that the expression of the variables $(\tl{p}_i,\tl{q}_i)_{i=1}^{n}$ is known explicitly in terms of the variables $(p_i,q_i)_{i=1}^{n}$. A \emph{sufficient} condition to obtain parametric and explicit BTs is to find a dressing matrix $D(\lm)$ that i) whose manifold coincides with a symplectic leaf of the same Poisson bracket as those satisfied by $L(\lm)$ \cite{S},  and ii) such that  its determinant has a non dynamical zero , i.e. $\det\left(D(\lm)\right)=0\Big|_{\lm=\mu}$, where $\mu$ is a parameter not depending on the dynamical variables. The condition i) ensures that the transformations are canonical (see \cite{S}), the condition ii) ensures that are explicit and parametric. Suppose indeed that the particular symplectic leaf defined by $D(\lm)$ is one dimensional, so that $D(\lm)$ will depend on another (up to now free) variable, say \emph{a}. Then, since $\det(D(\mu))=0$, $D(\mu)$ will have a kernel, say $|\Omega\rangle$, that in general also will depend on $\mu$ and \emph{a}. But from \ref{LD=DL} we see also that $|\Omega\rangle$ is an eigenvector of $L(\mu)$; indeed:
\be\label{kernel}
\tl{L}(\mu)\Big(\underbrace{D(\mu)|\Omega(\mu)\rangle}_{=0}\Big)=D(\mu)\Big(L(\mu)|\Omega(\mu)\rangle\Big) \Rightarrow
L(\mu)|\Omega(\mu)\rangle = \gm(\mu)|\Omega(\mu)\rangle 
\ee      
The last equivalence can be read as an equation for $a$ that, if solved, will give to $a$ a dependence on $\mu$ and on the dynamical variables of the system (only one set of them, in this case the ``untilded one''). So, returning to \ref{LD=DL}, one obtains parametric ($\mu$) and explicit transformations. \newline
Depending on the dimension of the leaf defined by $D(\lm)$ it is possible to obtain multi-parametric BTs. Also, one can obtain multi-parametric BTs through  repeated iterations of one parameter transformations, since BTs for finite dimensional systems always commute \cite{S}. \newline
For simplicity in the rest of the paper we take $L(\lm)$ to be a $\mathfrak{sl}(2)$ matrix, also if our results can be easily  generalized to $L(\lm) \in \mathfrak{gl}(2)$. For BTs found through a $N\textrm{x}N$ Darboux matrix see \cite{NRK}, \cite{S}. Please note that in the case $L(\lm) \in \mathfrak{sl}(2)$ the function $\gm(\mu)$ in \ref{kernel} satisfies $\gm^2(\mu)+\det(L(\mu))=0$ and is the generating function of the conserved quantities of the system.\newline
An additional property of BTs was introduced in \cite{KS}, namely the ``spectrality property''. In order to understand
this property remember that BTs are canonical transformations, implying that it exists a generating function $\FF_{1}(\tl{q},q,\mu)$ such that:
$$
p_i=\frac{\partial \FF_{1}}{\partial q_i}\qquad \tl{p}_i=-\frac{\partial \FF_{1}}{\partial \tl{q}_i}
$$
The spectrality property \cite{KS} says  that there exists a function $f_1$ satisfying the following equation:
\begin{equation}\label{spectr}
\frac{\partial \FF_{1}(\tl{q},q,\mu)}{\partial \mu}= -f_1(\gm(\mu),\mu)
\end{equation} 
where $\gm(\mu)$ is just the function in \ref{kernel}: it satisfies the relation $\det(L(\mu)-v(\mu)\dblone)=0$, that can be seen as the \emph{separation equation}, in the sense of Hamilton-Jacobi separability, for the dynamical system; by a classical point of view one obtains the quadrature of the equations, by a quantum point of view the factorization of the eigenfunctions \cite{S1} \cite{KS}, \cite{KV}. \newline
Starting by this point we show how the spectrality property can lead to the ``linearization'' of the maps defined by the BTs, that, in turns, give the general solution of the equations of motion. This construction give also us the possibilty to properly interpret the Baxter $Q$ operator (representing the quantum BTs) as the Green's function of the Schr\"odinger equation defined by the Hamiltonian interpolating the discrete flow given by the BTs.

\section{Spectrality property and separation of variables.}\label{Ch.1}
First of all we want to make some comments about the equation \ref{spectr} defining the spectrality property.
Suppose that it is possible to choose the parameter $\mu$ in such a manner that the dressing matrix $D(\lm, \mu)$ is proportional to the identity when $\mu=0$. Expanding equation \ref{LD=DL} around $\mu=0$, if $D(\lm,\mu)\doteq k(\dblone+\mu D_0(\lm)+O(\mu^2))$, one obtains:
\begin{equation}\label{limitflow}
\tl{L}(\lm)=L(\lm)+\mu[D_{0}(\lm),L(\lm)]+O(\mu^2)
\end{equation}
so that in the limit $\mu\to 0$, by defining $\dot{L}\doteq \lim_{\mu\to 0}\frac{\tl{L}-L}{\mu}$, the BTs define an Hamiltonian flow: in this sense $\mu$ can be considered an evolution parameter. Equation \ref{spectr} than resembles the Hamilton-Jacobi equation with respect to the time $\mu$, since the function $\gm(\mu)$ contains all the conserved quantities of the system. This point will be made clearer in the next lines.\\
Let us assume to have a set of BTs, with the parameter $\mu$ playing the role of a time, in the sense given before:
\begin{equation}\label{pqexpl} \left\{\begin{aligned}
&\tl{p}_i=\tl{p}_i(p_k,q_k,\mu)\quad \textrm{with}\quad \tl{p}_i\big|_{\mu=0}=p_i\\
&\tl{q}_i=\tl{q}_i(p_k,q_k,\mu) \quad \textrm{with}\quad \tl{q}_i\big|_{\mu=0}=q_i
\end{aligned}\right.
\end{equation}
These transformations can be also rewritten as (assuming obviously the implicit function theorem can be applied):   
\begin{equation}\label{pqimpl} \left\{\begin{aligned}
&\tl{p}_i=\tl{p}_i(\tl{q}_k,q_k,\mu)\\
&p_i=p_i(\tl{q}_k,q_k,\mu)
\end{aligned}\right.
\end{equation}
Since \ref{pqexpl} and \ref{pqimpl} are canonical transformations, there exists the respective generating functions, say $\mathcal{F}_{0}(p,q,\mu)$ and $\mathcal{F}_{1}(\tl{q},q,\mu)$, solving the corresponding system of differential equations:
\begin{equation}\label{F0F1} \left\{\begin{aligned}
&p_i-\sum_{k=1}^n\tl{p}_k\frac{\partial \tl{q}_k}{\partial q_i}=\frac{\partial \mathcal{F}_0}{\partial q_{i}}\\
&\sum_{k=1}^n\tl{p}_k\frac{\partial \tl{q}_k}{\partial p_i}=-\frac{\partial \mathcal{F}_0}{\partial p_{i}}
\end{aligned}\right.\qquad \qquad \left\{\begin{aligned}
&p_i=\frac{\partial \mathcal{F}_1}{\partial q_{i}}\\
&\tl{p}_i=-\frac{\partial \mathcal{F}_1}{\partial \tl{q}_{i}}
\end{aligned}\right.
\end{equation}
Now we assume that the transformations possess the spectrality property both with respect to $\mathcal{F}_{0}(p,q,\mu)$ and with respect to $\mathcal{F}_{1}(\tl{q},q,\mu)$. So there exist two functions, say $f$ and $g$, such that:
\begin{equation}\label{fg}
\frac{\partial \FF_0}{\partial\mu}=-f(\gm(\mu),\mu)\qquad \qquad
\frac{\partial \FF_1}{\partial\mu}=-g(\gm(\mu),\mu)
\end{equation}
This assumption has non trivial consequences as will be shown in the following two theorems. The result will be that, under the assumption, the BTs can be re-parametrized so to represent the solution of the Hamilton-Jacobi equation for the Hamiltonian interpolating the flow defined by \ref{limitflow}.      
\begin{th1}\label{th1}
Suppose that the spectrality property holds true both for $\mathcal{F}_{1}$ and $\FF_0$. Then 
$\sum_{k=1}^{n}\tl{p}_k\frac{\partial \tl{q}_k}{\partial \mu}$ is given by $\frac{\partial \FF_1}
{\partial\mu}-\frac{\partial \FF_0}{\partial\mu}$, that is $\sum_{k=1}^{n}\tl{p}_k\frac{\partial \tl{q}_k}{\partial 
\mu}$ is a function of only $\gm(\mu)$ and $\mu$.
\end{th1}
\textbf{Proof.} 
The two generating functions $\FF_0$ and $\FF_1$ are related by \cite{FM}:
\begin{equation}\label{eq1}
\FF_0(p,q,\mu)=\FF_1(\tl{q}(p,q,\mu),q,\mu)
\end{equation}
So, utilizing equation \ref{eq1} and $\tl{p}_i=-\frac{\partial \FF_1}{\partial \tl{q}_i}$ it is possible to write:
$$
\frac{\partial \FF_0}{\partial \mu}=\sum_{k=1}^n\frac{\partial \FF_1}{\partial \tl{q}_k}\frac{\partial \tl{q}_k}{\partial \mu}+\frac{\partial \FF_1}{\partial \mu}\Big|_{\tl{q}=const.}\Rightarrow \sum_{k=1}^n\tl{p}_k\frac{\partial \tl{q}_k}{\partial \mu}=f(\gm,\mu)-g(\gm,\mu) 
$$
\hspace*{140mm} $\Box$
\begin{th1}\label{th2}
Suppose that the assumption of \textrm{Theorem} \ref{th1} holds, then giving to $\mu$ a dependence on the constants of motion through any of the root(s) of the equation $\mu=h(g(\mu,\gm(\mu)))$, where $h$ is an arbitrary function, one obtains again a set of BTs. The generating function of the new transformations is given by:
\begin{equation}\label{eqth2}
\FF(p,q,\mu^{(k)})=\FF_0(p,q,\mu^{(k)})+\int^{\mu^{(k)}} g(\gm(\mu),\mu)d\mu \qquad
\end{equation}
where $\mu^{(k)}=h(g(\mu^{(k)},\gm(\mu^{(k)})))$ is any of the root of $\mu=h(g(\mu,\gm(\mu)))$
\end{th1}
\textbf{Proof.} 
Suppose to give to $\mu$ a dependence to the dynamical variables $(p_i, q_i)_{i=1}^{n}$. We ask if it is possible to choose some function $\mu(p,q)$ in such a way that the transformations $\tl{p}_{i}=\tl{p}_{i}(p,q,\mu(p,q))$ and $\tl{q}_{i}=\tl{q}_{i}(p,q,\mu(p,q))$ obtained inserting the function $\mu(p,q)$ in the expressions \ref{pqimpl} are again canonical. \\
If the new transformations are canonical they possess a generating function, say $\FF$, such that: 
\begin{equation}\label{F} \left\{\begin{aligned}
&p_i-\sum_{k=1}^n\tl{p}_k\frac{\partial \tl{q}_k}{\partial q_i}=\frac{\partial \mathcal{F}}{\partial q_{i}}\\
&\sum_{k=1}^n\tl{p}_k\frac{\partial \tl{q}_k}{\partial p_i}=-\frac{\partial \mathcal{F}}{\partial p_{i}}
\end{aligned}\right.
\end{equation}
Inserting the ansatz $\FF=\FF_0+A(p,q)$ into the previous system
and taking into account that by Theorem \ref{th1} $g=-\frac{\partial \FF_0}{\partial \mu}-\sum_{k=1}^n \tl{p}_k\frac{\partial \tl{q}_k}{\partial \mu}$, one readily finds the equations that has to be satisfied by $A(p,q)$:
\begin{equation}\label{Agmu} \left\{\begin{aligned}
&\frac{\partial A}{\partial q_i}=g\frac{\partial \mu}{\partial q_i}\qquad i=1..n\\
&\frac{\partial A}{\partial p_i}=g\frac{\partial \mu}{\partial p_i}\qquad i=1..n
\end{aligned}\right.
\end{equation}
Indeed from $\FF=\FF_0+A$ and from \ref{F} we have:
\begin{equation}\label{FF0A}
\frac{\partial \FF}{\partial p_i}=\frac{\partial\FF_0}{\partial p_i}\Big|_{\mu=const.}+\frac{\partial\FF_0}{\partial \mu}\frac{\partial\mu}{\partial p_i}+\frac{\partial A}{\partial p_i}=-\sum_{k=1}^{n}\tl{p}_k\frac{\partial \tl{q}_k}{\partial p_i}\Big|_{\mu=const.}-\sum_{k=1}^{n}\tl{p}_k\frac{\partial\tl{q}_k}{\partial \mu}\frac{\partial\mu}{\partial p_i}
\end{equation}
When $\mu$ is constant the BTs are canonical transformations with generating function $\FF_0$, so it holds the equation: 
$$
\frac{\partial\FF_0}{\partial p_i}\Big|_{\mu=const.}=-\sum_{k=1}^{n}\tl{p}_k\frac{\partial \tl{q}_k}{\partial p_i}\Big|_{\mu=const.}
$$ 
Inserting this equivalence in \ref{FF0A}, we are left with:
$$
\left(\frac{\partial \FF_0}{\partial \mu}+\sum_{k=1}^{n}\tl{p}_k\frac{\partial \tl{q}_k}{\partial\mu}\right)\frac{\partial \mu}{\partial p_i}+\frac{\partial A}{\partial p_i}=0
$$
But from Theorem \ref{th1} we have that $\frac{\partial \FF_0}{\partial \mu}+\sum_{k=1}^{n}\tl{p}_k\frac{\partial \tl{q}_k}{\partial\mu}=-g$, so finally:
$$
\frac{\partial A}{\partial p_i}=g\frac{\partial \mu}{\partial p_i} \qquad i=1..n
$$
In the same manner, by the equation for $\frac{\partial \FF}{\partial q_i}$ it is possible to show that: 
$$
\frac{\partial A}{\partial q_i}=g\frac{\partial \mu}{\partial q_i} \qquad i=1..n
$$
The compatibility equations for the function $A(p,q)$, that is $\frac{\partial^2 A}{\partial p_k\partial q_i}=\frac{\partial^2 A}{\partial q_i\partial p_k}$ give (by \ref{Agmu}):
\begin{equation}\label{gmu}
\frac{\partial g}{\partial p_k}\frac{\partial \mu}{\partial q_i}=\frac{\partial g}{\partial q_i}\frac{\partial \mu}{\partial p_k}\qquad i,k=1..n
\end{equation}
These equations means that $\mu$ has to be an arbitrary function of $g$, that is: 
$$
\mu=h(g(\mu,\gm(\mu)))
$$ 
Note that this (set of) implicit equation for $\mu$ fixes the dependence of $\mu$ by the dynamical variables $(p_i,q_i)_ {i=1}^{n}$ through its roots. At any of the root of this (set of) equation will correspond a function $g$ and a function $\mu$, both depending only on the dynamical variables $(p_i,q_i)_{i=1}^{n}$, satisfying the set of differential equations \ref{gmu}. Indicating the $k^{th}$ root of $\mu=h(g(\mu,\gm(\mu)))$ as $\mu^{(k)}$ and returning to the system \ref{Agmu}, one finds:
\begin{equation}\label{A}
A(p,q)=\int^{\mu^{(k)}}g(\gm(\mu),\mu)d\mu 
\end{equation}  
\hspace*{140mm} $\Box$\\
Let us now make some remarks that will be useful in what follows. 
\begin{remark}\label{rmrk1}
The generating function $\FF (p,q,\mu)$ as a function of $\mu$ satisfies the equation:
$$
\frac{\partial \FF(p,q,\mu)}{\partial \mu}=-\sum_{k=1}^{n}\tl{p}\frac{\partial \tl{q}}{\partial \mu}
$$
\end{remark}
This is just a trivial calculation on $\frac{\partial \FF(p,q,\mu)}{\partial \mu}$  for the function $\FF(p,q,\mu)$ as given in \ref{eqth2} and taking into account the Theorem \ref{th1}.
\begin{remark}\label{rmrk2} 
By adding a dependence on an extra constant parameter, say $T$, to the function $\mu$, that is by considering $\mu=\mu^{(k)}(T)$, the canonicity of the transformations is preserved, so again one has a set of BTs.
\end{remark}
This remark gives the possibility to obtain a \emph{new} parametric family of BTs by letting to $\mu^{(k)}$ to depend also by the parameter $T$. Indeed it is possible to repeat the same line of reasoning of the Theorem \ref{th2} by considering, from the beginning, $\mu$ as a function of the dynamical variables $(p_i,q_i)_{i=1}^{n}$ and the new constant parameter $T$. Obviously now also the generating function $\FF$ will got a dependence on the parameter $T$.\\
The key point of all this construction is that, assuming the spectrality property holds both for $\FF_0$ and $\FF_1$, it is possible, at least in principle, to obtain a new, larger, family of parametric BTs by giving to $\mu$ a dependence on the constants of motion and on the parameter $T$: as we will show now, to this freedom to obtain new BTs it corresponds the possibility to get, with a suitable choice of the function $\mu(p,q,T)$, the canonical transformation from the variables $(p_i,q_i)_{i=1}^{n}$ to the new variables $(\tl{p}_i,\tl{q}_i)_{i=1}^{n}$ such that the Hamilton-Jacobi equation, where the parameter $T$ plays the role of ``time'' and the interpolating Hamiltonian the role of its conjugated variable, is identically solved. \\
Summarizing, given a parametric BTs $(\tl{p}_i(p_k,q_k,\mu),\tl{p}_i(p_k,q_k,\mu))$ having the spectrality property both for $\FF_0$ and $\FF_1$, the goal would be to find a function $\mu$ fitting the Hamiltonian-Jacobi equation for the flux described by the BTs:
\begin{equation*}
\mathcal{H}+\frac{\partial F}{\partial \mu}\frac{\partial \mu}{\partial T}=0
\end{equation*}
where with the italic $\mathcal{H}$ we mean the interpolating Hamiltonian of the BTs. By Remark \ref{rmrk1} the previous equation can be better rewritten as:
\begin{equation}\label{H-J}
\mathcal{H}-\frac{\partial \mu}{\partial T}\sum_{k=1}^{n}\tl{p}\frac{\partial \tl{q}}{\partial \mu}=0
\end{equation}
Indeed now by Theorem \ref{th1} the sum $\sum_{k=1}^{n}\tl{p}\frac{\partial \tl{q}}{\partial \mu}$ is a function of $\mu$ and of the dynamical variables only through the constants of motion; obviously also the interpolating Hamiltonian $\mathcal{H}$ will depend only on some combinations of the constant of motions. So the equation \ref{H-J} is indeed solvable in terms of $\mu$. Inserting the solution $\mu(p,q,T)$ into the expressions of BTs we obtain, by construction, the \emph{general solution} of the equations of motion with respect the interpolating Hamiltonian $\mathcal{H}$.\\
Let us make another remark. It seems that there is an ambiguity in the choice of the function $\mu$ solving equation \ref{H-J} because, of course, the solution depends also on an arbitrary function of the constants of motion. But now we will show that the freedom to choose the value of this arbitrary function corresponds to the freedom in the choice of the initial conditions in the general solution of our equations of motion: so it represents a shift in the time $T$. Indeed, taking into account Theorem \ref{th1} and the equations \ref{fg}, the solution of the Hamilton-Jacobi equation \ref{H-J} can be implicitly written as: 
\begin{equation}\label{Tmu}
T+\frac{1}{\mathcal{H}}\int_{0}^{\mu}\left(g(\gm(\eta),\eta)-f(\gm(\eta),\eta)\right)d\eta + F(H_i)=0
\end{equation}    
where, by \ref{fg} we recall that the functions $f$ and $g$ are respectively given by $-\frac{\partial \FF_0}{\partial\mu}$ and $-\frac{\partial \FF_1}{\partial\mu}$ and $F$ is an arbitrary function of the Hamiltonians of the system $H_i$. It is now evident from equation \ref{Tmu} that i) the arbitrary function $F$ appears as an additive quantity with respect to the time parameter $T$ and ii) if one wants to retain the initial conditions as given in \ref{pqexpl}, that is $\tl{p}_i\big|_{T=0}=p_i$ and $\tl{q}_i\big|_{T=0}=q_i$, then one has to choose $F=0$.\\
In the next section we will give a simple application of the theorems just seen with the BTs for the one dimensional harmonic oscillator.

\section{A tutorial example: the harmonic oscillator}\label{Ch.2}
A Lax representation with spectral parameter for the one dimensional harmonic oscillator is given by:
\begin{equation*}
L(\lm) = \left(\begin{array}{cc} 1 & \frac{p-\textrm{i}q}{\lm}\\ \frac{p+\textrm{i}q}{\lm} & -1\end{array} 
\right) , \quad M = \frac{\textrm{i}}{2}\left(\begin{array}{cc} 1 & 0\\ 0& -1\end{array} \right), \quad \dot{L}(\lm)=[L,M]
\end{equation*}
In this simple case the spectral curve is simply given by:
$$
\gm^2(\lm)=-\det(L(\lm))=1+\frac{p^2+q^2}{\lm^2}
$$
In the following for brevity we pose $a\doteq p-\textrm{i} q$ and $a^*\doteq p+\textrm{i} q$.\\  
A class of BTs can be obtained by a \emph{dressing matrix} $D(\lm)$ parametrized as follows:
\begin{equation*}
D(\lm) = \left(\begin{array}{cc} 1 & \frac{\alpha}{\lm} \\ \frac{\beta}{\lm}& 1\end{array} 
\right) 
\end{equation*}
As pointed out in the introduction, to find \emph{explicit} transformations we impose that $\det(D(\lm=\mu))=0$. This constrain means that $\alpha\beta=\mu^2$, so we can rewrite, posing $\beta=\mu\zeta$:
\begin{equation*}
D(\lm) = \left(\begin{array}{cc} 1 & \frac{\mu}{\lm \zeta} \\ \frac{\mu \zeta}{\lm}& 1\end{array} 
\right) 
\end{equation*}
Obviously when $\lm=\mu$ the dressing matrix possesses a kernel $|\Omega(\mu)\rangle$. Explicitly it is given by:  
$$
|\Omega(\mu)\rangle=\left(\begin{array}{c}1\\-\zeta\end{array}\right)
$$ 
Then the eigenvector relation: 
$$
L(\mu)|\Omega(\mu)\rangle = \gm(\mu)|\Omega(\mu)\rangle 
$$
gives $\zeta$ as a function of $a$ and $a^*$:
$$
\zeta=\frac{\mu (1-\gm(\mu))}{a}=-\frac{a^*}{\mu(1+\gm(\mu))}
$$
Now we are able to write out the explicit form of BTs:
\begin{equation*}\label{explBTs} \left\{\begin{aligned}
&\tl{a}=a \frac{\gm(\mu)+1}{\gm(\mu)-1}\\
&\tl{a}^*=a^* \frac{\gm(\mu)-1}{\gm(\mu)+1}
\end{aligned}\right. \quad \textrm{with}\quad \gm(\mu)^2 = 1+\frac{|a|^2}{\mu^2}
\end{equation*} 
Note that when $\mu$ is equal to zero the transformations reduce to the identity map.\\
What we want to check now is whether these canonical transformations possess the spectrality property both for $\FF_0$ and for $\FF_1$. The generating function in the variables $a, a^*$ is given by $\FF_0(a,a^*,\mu)=-2\mu^2(\gm(\mu)+1)$, while the inverse transformations:
\begin{equation*} \left\{\begin{aligned}
&\tl{a} = \frac{4\mu^2 a^*}{(\tl{a}^*-a^*)^2}\\
&a = \frac{4\mu^2 \tl{a}^*}{(\tl{a}^*-a^*)^2}
\end{aligned}\right. 
\end{equation*}
are generated by the function $\FF_1(a^*,\tl{a}^*,\mu)=\frac{4\mu^2 a^*}{\tl{a}^*-a^*}$. 
By differentiating $\FF_0$ and $\FF_1$ one finds:
\begin{equation*} \left\{\begin{aligned}
&\frac{\partial \FF_0}{\partial \mu}=-2\mu\left(\sqrt{\gm(\mu)}+\frac{1}{\sqrt{\gm(\mu)}}\right)^2\\
& \frac{\partial \FF_1}{\partial \mu}\Big|_{\tl{a}^*=a^* \frac{\gm(\mu)-1}{\gm(\mu)+1}}=-4\mu(\gm(\mu)+1)
\end{aligned}\right. 
\end{equation*}  
so that both $\frac{\partial \FF_0}{\partial \mu}$ and $\frac{\partial \FF_0}{\partial \mu}$ are only functions of the integral ($p^2+q^2$) and of the parameter $\mu$. The results of the previous section can be applied. In the one dimensional case the Theorem \ref{th2} simply implies that $\mu$ can get an arbitrary dependence on the Hamiltonian and the BTs are again canonical transformations. So by now we consider $\mu$ as a function of $|a|$ and of a new parameter $T$, that is $\mu=\mu(|a|,T)$. Again from Theorem \ref{th2} the new transformations are again canonical with the generating function:  
$$
\FF=\FF_0\Big|_{\mu=\mu(|a|)}+4\int^{|a|}\mu(\gm(\mu)+1)\frac{\partial\mu(x,T)}{\partial x}dx
$$
In order to fix the dependence of $\mu$ by $|a|$ and $T$ we need to find the interpolating Hamiltonian. As explained at the beginning of the section \ref{Ch.1} the interpolating Hamiltonian of the flow with respect to $\mu$ can be extrapolated  considering the relation $\tl{L}D=DL$ in the limit $\mu \to 0$:
\begin{equation}\label{gty}
D=\dblone-\mu D_{0}+O(\mu^2)\Rightarrow \dot{L}(\lm)\doteq \lim_{\mu\to 0}\frac{\tl{L}-L}{\mu}=\left[L(\lm),D_0(\lm)\right]
\end{equation} 
In our case the matrix $D_{0}(\lm)$ is given by:
$$
D_0(\lm) = \left(\begin{array}{cc} 0 & \frac{1}{\lm}\sqrt{\frac{a}{a^*}}\\ \frac{1}{\lm}\sqrt{\frac{a^*}{a}} & 0\end{array} 
\right) 
$$
The continuous flow given by \ref{gty} is governed by the Hamiltonian $\mathcal{H}=2\textrm{i}|a|$. Note that the factor ``i'' appear because $a$ and $a^*$ are not canonically conjugate. \\
Now we have all the ingredients of the recipe. So we can look at the Hamilton-Jacobi equation:
\begin{equation}\label{muTi}\begin{split}
& \mathcal{H}-\tl{a} \frac{\partial \tl{a}^*}{\partial \mu}\frac{\partial\mu}{\partial T}=0 \rightarrow 2\textrm{i}|a|+\frac{2|a|^2}{\sqrt{\mu^2+|a|^2}}\frac{\partial\mu}{\partial T}=0 \rightarrow \\
& \rightarrow \mu(|a|,T)=-\textrm{i}|a|\sin(\frac{T}{|a|}+F(|a|))
\end{split}\end{equation}
As noted at the end of section \ref{Ch.1}, one can choose the arbitrary function $F(|a|)$ to be zero by requiring that $\mu\Big|_{T=0}=0$. \\
Inserting the formula $\mu(|a|,T)=-\textrm{i}|a|\sin(\frac{T}{|a|})$ into the expressions of BTs \ref{explBTs} and returning to the physical variables $p=\frac{a^*+a}{2}$ and $q=\frac{a^*-a}{2\textrm{i}}$ one finds:
\begin{equation*} \left\{\begin{aligned}
&\tl{p}=p\cos(\frac{2T}{\sqrt{p^2+q^2}})-q\sin(\frac{2T}{\sqrt{p^2+q^2}})\\
&\tl{q}=q\cos(\frac{2T}{\sqrt{p^2+q^2}})+p\sin(\frac{2T}{\sqrt{p^2+q^2}})
\end{aligned}\right. 
\end{equation*}
that is the general solution of the equations of motion governed by the Hamiltonian $\mathcal{H}=2\sqrt{p^2+q^2}$. Note that we now omit the factor ``i'' from the interpolating Hamiltonian because we are considering directly the flow with respect to $T$ and not with respect to $\mu$  (when $T$ goes to zero $\mu \sim -\textrm{i}T$).\\
Note that the parameter $T$ is a ``linearizing parameter'' for the flow.\\
For completeness we have to cite \cite{RZK}, where another way to find the functional relation between $\mu$ and $T$ (if any!) has been given. However in \cite{RZK} there were no definite statement about the necessary and sufficient condition for such relation to exist, neither its connections with analytic mechanics was pointed out. Nevertheless that point of view can be very useful because to the explicit knowledge of the BTs does not always correspond the explicit knowledge of the generating functions $\FF_0$ and $\FF_1$ of the transformations. It can be also shown that applying the machinery described in \cite{RZK}, one obtains exactly the result $\mu(|a|,T)=-\textrm{i}|a|\sin(\frac{T}{|a|})$, as in \ref{muTi}. The reader is referred to this work for other examples of linearization of BTs.

\section{Quantum B\"acklund transformations: some hints and insights.}\label{Ch.3}
By a classical point of view the BTs associate new solutions of the equations of motion to a given one. So, in general, if $g(p(t),q(t))$ is a physical observable connected to the curve in the phase space $(p(t), q(t))$, $g(\tl{p}(t),\tl{q}(t))$ will be the same physical observable connected to the curve$(\tl{p}(t),\tl{q}(t))$. \\
By a quantum point of view the BTs are represented by a unitary operator $Q^{(1)}_\mu$ ($\mu$ being the (set of) parameter(s) of the BTs) realized as an integral operator on the space of eigenfunctions \cite{PG}, \cite{KS}:
\begin{equation}\label{Qf}
Q^{(1)}_\mu :\psi(q) \to \int f^{(1)}(\tl{q},q)\psi(q)dq
\end{equation}
The similarity transformations induced by $Q^{(1)}_\mu$ are the equivalent of the classical canonical transformations and the kernel $f^{(1)}(\tl{q},q)$ is given, in the semiclassical approximation, by:
\begin{equation}\label{SC}
f^{(1)}(\tl{q},q) \sim \exp(-\frac{\textrm{i}}{\hbar}F_1(\tl{q},q)), \qquad \hbar \to 0 
\end{equation}
As far as we know, the explicit construction of the $Q$ operator, known in literature as \emph{Baxter operator}, \emph{and} its relations with BTs, have been pointed out only for the Toda lattice \cite{PG}, \cite{S} and for the discrete self-trapping (DST) model \cite{KSS}. 
By our point of view, when the BTs can lead to the integration of the equations through the construction given in the previous sections, the kernel $f^{(1)}(\tl{q},q)$ has to be identified with the propagator or the Green's function for the time dependent Schr\"odinger equation, the Hamiltonian being that interpolating the flow represented by the BTs. Indeed, in this case, to the classical time shift on the trajectory specified by the initial conditions represented by the BTs it corresponds, as it is clear also from eq. \ref{Qf}, a quantum probability amplitude for the particle with position eigenvalue $q$ at the time $0$ to be found at a later time $\mu$ in $\tl{q}$. \\
In the following we make a check on the Baxter operator for the harmonic oscillator by showing that indeed its kernel gives exactly the well-known propagator of the corresponding Schr\"odinger equation. In order to compare the results 
we make a slight modification in the choice of the function $\mu$ in \ref{muTi}, by posing $T=\frac{\phi |a|}{2}$. Now $\phi$ play the role of the time. With this choice indeed the interpolating Hamiltonian is just the physical one, and obviously the BTs give the formulae:  
\begin{equation*}\left\{\begin{aligned}
&\tl{p}=p\cos(\phi)+q\sin(\phi)\\
&\tl{q}=q\cos(\phi)-p\sin(\phi)
\end{aligned}\right.
\end{equation*}
In order to find the generating function $F_{1}(\tl{q},q)$ we express $\tl{p}$ and $p$ in terms of $q$ and $\tl{q}$: 
\begin{equation*} \left\{\begin{aligned}
&\tl{p}=\frac{q-\tl{q}\cos(\phi)}{\sin(\phi)}\\
&p=\frac{q\cos(\phi)-\tl{q}}{\sin(\phi)}
\end{aligned}\right. 
\end{equation*}
From $dF_1=pdq-\tl{p}d\tl{q}$ one readily finds:
$$
F_1(\tl{q},q)=\frac{(\tl{q}^2+q^2)\cos(\phi)-2q\tl{q}}{2\sin(\phi)}+w(\phi)
$$
where $w(\phi)$ is an arbitrary function of $\phi$.\\
Now we have to find the $Q$ operator, or, that is the same, its kernel $f^{1}(\tl{q},q)$.  In the $\langle\textrm{bra}|\textrm{ket}\rangle$ notation the kernel $f^{1}(\tl{q},q)$ is given, in the $q$ representation, by $\langle \tl{q}|q\rangle$. So, by the following form of the BTs:
\begin{equation*} \left\{\begin{aligned}
&q=\tl{q}\cos(\phi)+\tl{p}\sin(\phi)\\
&\tl{q}=q\cos(\phi)-p\sin(\phi)
\end{aligned}\right. 
\end{equation*}
it is easily found that $f^1(\tl{q},q)$ solves the system:
\begin{equation*} \left\{\begin{aligned}
&qf^{(1)}=\tl{q}f^{(1)}\cos(\phi)-\textrm{i}\hbar\frac{\partial f^{(1)}}{\partial \tl{q}}\sin(\phi)\\
&\tl{q}f^{*(1)}=qf^{*(1)}\cos(\phi)+\textrm{i}\hbar\frac{\partial f^{*(1)}}{\partial \tl{q}}\sin(\phi)
\end{aligned}\right.
\end{equation*}
where $^{\centerdot^*}$ means complex conjugation. The solution of the previous system is:
$$
f^{(1)}(\tl{q},q)=C(\phi)\exp\left(-\frac{\textrm{i}}{\hbar}\left(\frac{(\tl{q}^2+q^2)\cos(\phi)-2q\tl{q}}{2\sin(\phi)}\right)\right)
$$
in exact agreement with \ref{SC}. \\
For completeness we recall the classical results on the propagator of the Schr\"odinger equation for the harmonic oscillator. In general the propagator can be formally written as:
$$
K(\tl{q},q)=\langle \tl{q}|\exp(-\frac{\textrm{i}H t}{\hbar})|q\rangle =\sum_{n}\langle \tl{q}|n\rangle\langle n|q\rangle\exp(-\frac{\textrm{i}E_n t}{\hbar})
$$
In the case of the harmonic oscillator $\langle q|n\rangle$ are the Hermite polynomials, so that:
$$
K(\tl{q},q)=\sum_{n}\frac{e^{-\left(\frac{\tl{q}^2+q^2}{2\hbar}\right)}}{2^n n!\sqrt{\pi\hbar}}H_n(\frac{\tl{q}}{\sqrt{\hbar}})H_n(\frac{q}{\sqrt{\hbar}})e^{-\textrm{i}\left(n+\frac{1}{2}\right)\phi}
$$
It is possible to sum up the series by using the so called "Meheler's formula" (see e.g. \cite{MF}):
\begin{equation}\label{meheler}
\sum_n \frac{w^n}{2^n n!}H_n(x)H_n(y)=\frac{1}{\sqrt{1-w^2}}\exp\left(\frac{2xyw-(x^2+y^2)w^2}{1-w^2}\right)
\end{equation}
obtaining:
$$
K(\tl{q},q)=\frac{1}{\sqrt{2\pi\textrm{i}\hbar\sin(\phi)}}\exp\left(-\frac{\textrm{i}}{\hbar}\left(\frac{(\tl{q}^2+q^2)\cos(\phi)-2q\tl{q}}{2\sin(\phi)}\right)\right)
$$
This formula agrees with that of $f^{(1)}(\tl{q},q)$ identifying $C(\phi)$ with $\frac{1}{\sqrt{2\pi\textrm{i}\hbar\sin(\phi)}}$.\\
On the same line of reasoning we can also ask for the quantum probability amplitude for the particle with momentum eigenvalue $p$ at the time $0$ to possess at a later time $\mu$ the momentum $\tl{p}$. Again there is the correspondence between the classical generating function, now given by $d\FF_2=\tl{q}d\tl{p}-qdp$, and the kernel of the corresponding Baxter operator $Q^{(2)}_\mu$:
\begin{equation}\label{SC2}
f^{(2)}(\tl{p},p) \sim \exp(-\frac{\textrm{i}}{\hbar}F_2(\tl{p},p)), \qquad \hbar \to 0 
\end{equation}
From the equations:
\begin{equation*} \left\{\begin{aligned}
&\tl{q}=\frac{\tl{p}\cos(\phi)-p}{\sin(\phi)}\\
&q=\frac{\tl{p}-p\cos(\phi)}{\sin(\phi)}
\end{aligned}\right. 
\end{equation*}
one readily finds:
$$
F_2(\tl{p},p)=\frac{(\tl{p}^2+p^2)\cos(\phi)-2p\tl{p}}{2\sin(\phi)}+\eta(\phi)
$$
where again $\eta$ is an arbitrary function. Whereas from the following form of the BTs:
\begin{equation*} \left\{\begin{aligned}
&\tl{p}=p\cos(\phi)+q\sin(\phi)\\
&p=\tl{p}\cos(\phi)-\tl{q}\sin(\phi)
\end{aligned}\right. 
\end{equation*}
one finds, recalling that in the $p$ representation $q=\textrm{i}\hbar\frac{\partial}{\partial p}$, the equations to be satisfied by $f^{(2)}$ as:
\begin{equation*} \left\{\begin{aligned}
&pf^{(2)}=\tl{p}f^{(2)}\cos(\phi)-\textrm{i}\hbar\frac{\partial f^{(2)}}{\partial \tl{p}}\sin(\phi)\\
&\tl{p}f^{*(2)}=pf^{*(2)}\cos(\phi)+\textrm{i}\hbar\frac{\partial f^{*(2)}}{\partial \tl{p}}\sin(\phi)
\end{aligned}\right.
\end{equation*} 
so that:
$$
f^{(2)}(\tl{p},p)=V(\phi)\exp\left(-\frac{\textrm{i}}{\hbar}\left(\frac{(\tl{p}^2+p^2)\cos(\phi)-2p\tl{p}}{2\sin(\phi)}\right)\right)
$$ 
again in exact agreement with \ref{SC2}.

\section{Conclusions}
In this work we tried to shed some light on the spectrality property of BTs, by showing how it can lead, through the constructions of the section \ref{Ch.2}, to the explicit integration of the underlying equations of motion. It could be interesting to apply these results to many-body systems possessing known BTs with spectrality property, such as, for example, the Toda lattice. Also the interpretation of the Baxter operator for such systems as the propagator for the corresponding Schr\"odinger operator could lead to interesting summation formulae for orthogonal polynomials, similar to the Meheler's formula \ref{meheler}. As a final remark we point out that the non-obvious relation among B\"acklund transformations and the Green's function of the Schr\"odinger operator provides a bridge between the theory of such transformations and the Feynman path integral approach, opening unexplored perspectives in statistical mechanics and quantum field theory.

\end{document}